\documentstyle[12pt, aasms4]{article} 

\def\ba       {{$b/a$ }}
\def\fba      {{$f(b/a)$ }}
\def\logR     {{$\log R_{25}$ }}
\def\logD     {{$\log D_{25}$ }}

\def\Uband    {{$U_{300}$} }
\def\Bband    {{$B_{450}$} }
\def\Vband    {{$V_{606}$} }
\def\Iband    {{$I_{814}$} }
\def\Bj       {{$B_{J}$} }
\def\Bccd     {{$B_{CCD}$} }
\def\Bt       {{$B_{T}$} }
\def\U        {{$U$} }
\def\B        {{$B$} }

\def\Vgsr     {{$V_{GSR}$} }

\def\I        {{$I$} }
\def\etal     {{et\thinspace al.} }
\def\cf       {{\it cf.} }
\def\eg       {{\it e.g.}, }
\def\ie       {{\it i.e.}, }

\def\cge      {{$_ >\atop{^\sim}$}}
\def\cle      {{$_ <\atop{^\sim}$}}
\def\degsq    {{\ $deg^{2}$} }

\def\Ho       {{$H_{0}$} }
\def\qo       {{$q_{0}$} }

\def\arcspt   {{$\buildrel{\prime\prime}\over .$}}

\def\magarc   {{\ mag\ arcsec$^{-2}$} }
\def\SN       {{S/N} }
\def\deg      {{\ifmmode^\circ\else$^\circ$\fi} } 

\font\ksmit=cmti10     scaled \magstephalf

\begin{document} 

\begin{center} 

\title{
The Axis Ratio Distribution of Local and Distant Galaxies. \footnote{Based on
observations with the NASA/ESA {\it Hubble Space Telescope} obtained at the
Space Telescope Science Institute, which is operated by AURA, Inc., under NASA
Contract NAS 5-26555.} }

\author{Stephen C. Odewahn \footnote{Currently at Department of Astronomy, California Institute of Technology}, David Burstein, and Rogier A. Windhorst} 

Dept. of Physics \& Astronomy, Arizona State University, Box 871504, Tempe,
AZ\ 85287-1504\

Email: sco@poseidon.caltech.edu, raw@cosmos.la.asu.edu,
burstein@samuri.la.asu.edu

\vspace{1.5cm} 
Send preprint requests to ``sco@poseidon.caltech.edu'' 
\end{center} 

\begin{center}
{\bf Date of this version: {\large \today} } 
\end{center}

\begin{abstract} 

Surface photometry from 16 HST/WFPC2 fields in the \Iband-band filter is used
to derive the distribution of apparent axis ratios for galaxies in
progressively fainter magnitude intervals for \Iband$\leq$25. 
We assess the systematic and
accidental errors in ellipticity measurements as a function of image
resolution and signal-to-noise ratio, and statistically correct for the effect
of cosmological surface brightness dimming on our isophotal measurements. The
axis ratio distribution for the local galaxy population was computed using
\logR measurements for 1569 RC3 galaxies with \Bt$\leq$13 mag. Nonparametric
tests are used to show that our distant samples, in the redshift range 0.1\cle
z \cle 1.0--1.5, are not statistically different from the local sample. We
present image montages of galaxies selected randomly from different axis ratio
and apparent magnitude ranges and discuss the evolutionary consequences of
the lack of a strong difference between the ellipticity distributions in 
near and far data sets.

\end{abstract} 

{\it Subject headings:} galaxies: elliptical --- galaxies: spiral ---
galaxies: irregular --- galaxies: evolution --- galaxies: fundamental
parameters


\section{INTRODUCTION} 
  
The apparent isophotal axis ratio, $b/a$, of a galaxy is related to its
intrinsic three-dimensional structure.  A statistical recovery of the
frequency distribution of intrinsic shapes using apparent axis ratios is
dependent on the condition that galaxies are randomly oriented in space with
respect to the observer.  It is also dependent on the condition that known
biases in the galaxy sample, such as those of Malmquist-type, or measurement
systematics, and dust, etc., can be adequately modeled.

The first detailed study of the axial ratios of galaxies was made by Hubble
(1926), who established the basic methodology used in later papers by de
Vaucouleurs (1959), Sandage, Freeman \& Stokes (1970), and de Vaucouleurs \&
Pence (1973).  Subsequent work by Noerdlinger (1979) and Binney \& de
Vaucouleurs (1981) used mathematical inversion techniques to transform the
observed \fba to the intrinsic distribution. Odewahn (1989) applied the
modeling methods of the earlier papers to the much improved axis ratios and
diameters that refer to a standard isophotal level in the RC3 (\cf de
Vaucouleurs \etal 1991) to study intrinsic galaxy shapes as a function of
morphological type.  More extensive studies of this type, employing larger
samples of photographic and CCD galaxy surface photometry are presented by
Ferguson \& Sandage (1989), Fasano \& Vio (1991), Lambas \etal (1992), and
Ryden \& Terndrup (1994). 


The principal aim of these papers was to determine the distribution of
intrinsic shapes for galaxies drawn from different population samples, \eg
from different morphological types or surface brightness classes. The
intrinsic shape of the optical component of a galaxy is likely to be linked to
the physical state of the dark matter halo and hence may present important
information about the history and dynamical state of the system.
Yet, any statistical approach to determine true axial ratios from observed
ellipticities is valid only if two simplifying assumptions are made, and
various systematic effects are taken into account.

These two simplifying assumptions are: a) that galaxies are
randomly oriented in space with respect to the observer; and b) that all
galaxies in a given sample have the same intrinsic three-dimensional form.
Systematic effects include surface-brightness selection effects,
Malmquist-bias types, sky subtraction errors, internal dust properties, 
and triaxiality of early-type (elliptical) galaxies. Hence,
in practice it is very difficult to derive the 3-D shapes of galaxies.
Lambas \etal (1992) find strong evidence for triaxiality
among the early-type ellipticals, a situation which complicates the derivation 
of intrinsic shapes using the apparent axis ratio distribution.
Our present purpose is {\it not} to derive the intrinsic shapes of galaxies as a
function of Hubble type.  Rather, we wish to use the observed ellipticity
distribution as tool to search for evidence of dynamical galaxy evolution over
a reasonably large fraction of the age of the Universe. We want to investigate
if different galaxy populations with different ellipticity distributions (\ie
disk- or bulge-dominated populations) are present at different flux levels
(\ie, different lookback times). In particular, we want to investigate if the
epoch of the onset of disk formation can be delineated from the axis-ratio
distribution of galaxies observed with the Hubble Space Telescope (HST).

Recently, Im \etal (1995) used the form of \fba to make inferences about the 
faint galaxy populations surveyed with HST. The $\sim$0\arcspt 1 FWHM resolution 
provided by HST's Wide Field Planetary Camera 2 (WFPC2) allows the determination 
of the sub-kpc morphology and isophotal shapes of distant galaxies --- as well as 
the population fractions of different galaxy types, over a wide range of epochs.
The morphological properties of the faint galaxy population have recently been
studied with HST by Driver \etal (1995a, D95a; 1995b, D95b), Glazebrook \etal
(1995), Cowie \etal (1995), Odewahn \etal (1996), and Abraham \etal (1996).
These morphology-based studies found that a wider range of peculiar or
disturbed morphologies are observed at faint magnitudes ( 20\cle $I$\cle
25mag, 0.5\cle $z$ \cle 1--2), compared to what is seen in the local Universe.
Uncertainties arising from cosmological effects such as bandpass shifting and
$(1+z)^{4}$ surface brightness (SB) dimming complicate the interpretation of
morphological classification systems for studying galaxy evolution (Burg 
\etal 1997).  The goal of the present study is therefore to compare the 
properties of nearby galaxies to those seen at high redshift with HST without 
regard to predicted Hubble type. 


These tests using the galaxy axial ratio distributions are a useful
counterpart to morphological studies to understand how galaxies at high $z$ 
compare to those near us. Im \etal (1995) presented the axis ratio
distributions for different HST galaxy types, as determined by bulge- and
disk-decomposition of the HST images. This, however, assumes that the galaxy
types can be determined reliably and independently from the axis ratios. It is
not a priori clear that this can be done, and that the axis ratios therefore
are independent of the galaxy type. Strictly speaking, we do not
know which galaxy types seen at high redshift evolve into which kind of local
counterparts in a universe if hierarchical clustering dominates, and 
therefore we prefer not to separate the axis ratios as function of galaxy type.
Therefore, in this paper, we will compute the axial distributions for {\it
all} galaxies detected on the deep HST images, irrespective of their predicted
Hubble type. While we forego quantitative information of the axial ratio
distributions of specific Hubble types, we benefit from inclusion of all
galaxies, whether or not easily classified in a morphological type based on
present-day galaxies (e.g., linear or ``chain'' galaxies).

In \S 2, we discuss the observations and their reduction using the automated
image analysis system MORPHO. In \S 3, we discuss the systematic and
accidental errors in axis ratio measurements in the regime of low
resolution and/or low \SN ratio. We also discuss the errors that exist in the axial
ratio distribution of the local galaxy sample from the RC3.  In \S 4 we
present our determination of the frequency distributions of apparent axis
ratio in progressively fainter magnitude intervals. In \S 5 we discuss how our
results can be related to existing morphological tests of galaxy evolution in
terms of the hierarchical clustering/merger paradigm for the formation of
galaxies.

\section{OBSERVATIONS and REDUCTION}

\subsection{Observations }

The HST data archive was accessed via STEIS to obtain a number of random WFPC2
fields with F814W images.  The F814W filter was chosen to make a consistent
measurement of axial ratios in the redshifted passband which --- 
for 0.5 \cle $z$ \cle 1.0 --- comes closest to the B passband for local
galaxies.  The deepest field analyzed in this work was the Hubble Deep Field
(HDF; Williams \etal 1996), which for the present purposes has 49 exposures in
F814W, yielding a $1\sigma$ surface brightness (SB) sensitivity of $\rm
I_{814} \le 28$ \magarc, and a 5$\sigma$ point-source sensitivity of
\Iband$\simeq$29.0 mag (Odewahn \etal 1996).

A deep WFPC2 field surrounding the weak radio galaxy 53W002 at $z=$2.39, was
imaged in Cycles 4--5 using the F814W filter for 12 orbits (see \eg Driver
\etal 1995a). These dithered images reach a $1\sigma$ SB sensitivity of $\rm
I_{814} \le 26.7$ \magarc, and a $5\sigma$ point source sensitivity of
\Iband$\simeq$28.0 mag (Windhorst \etal 1997).  Moreover,
the dark HST sky at 53W002's location near the North Ecliptic Pole and the
small galaxy scale-lengths (Odewahn \etal 1996) allow detection of compact
galaxies in this field down to $I$\cle 27 mag.  Morphological studies of these
images (hereafter referred to as the $``$W02'' data set) are presented by Driver
\etal (1995a) and Odewahn \etal (1996).

Two other sets of observations go less deep than HDF and W02: (1) the eight
WFPC2 flanking fields surrounding the HDF (Williams \etal 1996) have mean
integration times from 2500 sec to 5300 sec (hereafter referred to as the
``HFF'' data set).  Six Medium Deep Survey Fields have mean integration times
from 866 sec to 2100 sec (Driver, Windhorst, \& Griffiths \etal 1995b; the
``DWG'' data set). Together, these 14 fields cover a sky area of 0.0168
\degsq. Based on their total F814W galaxy number counts, we estimate that
these shallower images have a 90\% completeness limit of \Iband\cle 24.0 mag
for galaxies of average SB, although our use of these fields, listed in 
Table~\ref{tab:TAB1}, will be restricted to the higher \SN images with 
total flux \Iband\cle22 mag.

\subsection{Data processing and catalog generation }

Initial photometric catalogs of galaxies in each WFPC2 field were made using
the SExtractor software (Bertin \& Arnouts, 1996).  This package is designed
to locate discrete galaxy images and extract photometric parameters for
extended sources in crowded galaxy fields. Candidates are defined to have 
at least 8 connected pixels with flux $\ge 2\sigma$ above the local median sky. 
In each case, a 3$\times$3 gaussian
smoothing filter is applied prior to detection. Adequate \SN is required to
determine reliable isophotal axis ratios, so we cut our final catalogs well
above the formal limiting magnitudes of each sample: \ie \Iband$\leq$22.0 mag
for the HFF and DWG fields, \Iband$\leq$25.0 mag for the W02 field, and
\Iband$\leq$26.0 mag for the HDF. Photometric zeropoints for each WFPC2 image
were computed from the mean WFPC2 integration times and the synthetic zero
points listed in Table 9 of Holtzmann \etal (1995).

Automated surface photometry of all cataloged objects was performed with the
software package MORPHO (Odewahn 1995,1997; Odewahn \etal 1996), designed for
neural network-based morphological classification of galaxies. The initial
galaxy catalogs generated by SExtractor are used to cut postage stamp images
around each detected galaxy. Local sky values are substituted for faint galaxy
images close to each given object in terms of elliptical areas whose size,
shape and orientation are established in a previous iteration by SExtractor.
Local sky estimates are made by the MORPHO package by iteratively fitting a
sloped plane to a surrounding empty sky-box (see Windhorst \etal 1991),
excluding the area of all other objects found by SExtractor in that sky-box.
We used for the first step an iterative rejection algorithm that determines
the best local sky mode.

We compared the sky values from this automated surface photometry to those for
$\sim$50 faint objects in the {\ksmit same} W02 images measured by Pascarelle
\etal (1996) with a completely independent {\ksmit interactive} package, that
otherwise similarly used a sloped plane-fit to a surrounding visually empty
sky-box. Both methods produce sky-estimates consistent within 0.07\% (and
consistent with Monte Carlo tests of the accuracy of this {\ksmit interactive}
sky determination by Windhorst \etal 1991; W91).  The two methods also give
total aperture magnitudes mutually consistent within 0.05 mag, with
rms$\simeq$0.22 mag for each algorithm at the faint flux levels. Hence, we 
believe that our sky-errors are dominated by random (fitting) errors, but
not by large-scale systematics. 

We also compare the total magnitudes from the MORPHO ($I_{S}$) and 
SExtractor ($I_{M}$) packages for about 1200 galaxies. 
Using galaxies with \Iband$\leq$22, a mean zeropoint
offset of $\langle I_{S}-I_{M} \rangle = -0.01$ was derived (negligibly
small). A 3\% scale error, significant at the 3$\sigma$ level, was evident
between the MORPHO and SExtractor magnitude systems. This
result was anticipated since we compare the ``Kron'' magnitudes of SExtractor
(magnitudes extrapolated to near-total using the second moment of the light
distribution) to the isophotal magnitudes of MORPHO (integrated magnitude
measured within an isophotal aperture sized to match a fiducial isophotal
level). As expected, the SExtractor magnitudes are consistently brighter than
the corresponding MORPHO values. For I$\leq$24 mag, this systematic difference
is described well by:
 
\begin{equation}
I_{S} - I_{M} =  -0.01 - 0.035(I_{S}-22)   \label{eq:eq1}
\end{equation}
 
We have adopted the MORPHO magnitudes for use in this
paper, but give the following relation to bring the isophotal
MORPHO magnitudes onto the total SExtractor system:
 
\begin{equation}
I_{S} =  0.966I_{M} + 0.73                 \label{eq:eq2}
\end{equation}
 
Over the range of magnitudes to be investigated here, 19$<$\Iband$<$25, this
systematic difference will result in a change of at most 0.25 magnitudes, or
12\% of the width of the binning interval used in this work to divide our
galaxy samples.
 
Surface photometry is performed on each of the patched postage stamp images
using the MORPHO image analysis package.  This provides ellipse fits over a
range of isophotal levels (based on the local sky-noise) and elliptically
averaged surface brightness profiles, as well as sets of photometric
parameters known to show significant dependence with morphological type for
more nearby galaxies as well as HST galaxies (Odewahn \& Aldering 1995,
Odewahn \etal 1996).

As discussed in \S 4, we expect that a significant portion of our faint galaxy
sample (I\cge24) will have redshifts of $z$\cge0.5, and hence will suffer
significant cosmological SB dimming. For this reason, an
ellipse fitting package was added to MORPHO that uses axis ratio measurements
over a range of isophotal levels and that uses an estimate of the redshift, $z$,
based on the expected median redshift vs. \Iband magnitude relation. For this,
we used the measured redshift-magnitude relation from Lilly \etal (1995) for 
$I\leq22$ mag, and photometric redshift estimates by Sawicki \etal (1996) for
$I\leq25$ mag, to estimate the proper {\it restframe} surface brightness level
for measuring the apparent axis ratio \ba.

We cleaned our final galaxy sample using an interactive MORPHO package which
plots mosaics of the selected galaxy postage stamps with the isophotal ellipse
fits overplotted. In some cases, image merging due to nearby galaxies or
diffraction spikes from nearby stars made it impossible to compute reasonable
ellipse fits. Such cases, comprising less than 5\% of the cataloged images,
were interactively rejected from subsequent analysis. These rejected galaxies 
were isolated and their \fba distribution was compared to that of the bonafide 
galaxy sample to assure that no strong signature in the apparent axis ratio 
distribution was being removed.

\section{SYSTEMATIC ERRORS IN AXIS RATIO MEASUREMENTS}

Given the fundamental differences between the HST images and those from
different ground-based telescopes, it is inevitable that systematic errors
will exist for any comparison of axial ratios of galaxies observed with these
instruments. As a result, we should expect that comparisons between photometric 
properties of local and high redshift galaxies will likewise suffer from systematic 
measurement errors. We note that such errors occur at both low
redshift and at high redshifts.  For example, even with the superb resolution
of HST, images of high redshift galaxies ($z$\cge1) suffer from poor linear
resolution and low \SN. As we go fainter with the HST images, our ability to
measure meaningful isophotal parameters such as size and shape becomes
progressively impaired. Separately, existing axial ratio data for nearby
galaxies suffer either from being a mix of direct measurements from images and
eye estimates (such as in the RC3), and/or from problems of zero point
calibration and sky subtraction issues (e.g., purely CCD-based surveys).

We adopt the point of view that there is no one ``correct'' measurement of the
axial ratios of galaxies, but rather a series of different kinds of
measurements that need to be brought onto the same internal system. As such,
it is imperative to understand how our apparent axis ratio measures are
affected by these systematics as we progress to fainter magnitude intervals. 
Similarly, we need to test how consistent are our ellipticity data for nearby
galaxies.

\subsection{Image Simulations of the Distant Galaxy Sample}

The MORPHO program allows us to simulate the appearance of a galaxy of known
redshift at some user-specified higher redshift. This package is designed
primarily to simulate the effects of cosmological SB dimming
at high redshift predicted by the equations given in Weedman (1986), 
with K-corrections of Lilly \etal (1995). With the $\Theta-z$ relation used by
Bohlin \etal (1991), Hill \etal (1992), and Mutz \etal (1994) for \Ho=75
km/s/Mpc and \qo=0.5, images of local galaxies with high resolution and \SN
were repixelated to yield the resolution that would have been observed at
redshift $z$ with the WFPC2 pixel size of 0\arcspt0996. Sets of high redshift
images were simulated using two sets of input images: (1) 63 \Bj images of
local galaxies ($z\leq0.05$) from the Frei \etal (1996) sample, and (2) a
subset of 223 galaxies observed with WFPC2 in \Iband with  \I$\leq$22 mag from
Driver \etal (1995a, 1995b). In both cases, we want to simulate the appearance
of galaxies observed at high redshift ($z$\cge0.5-1) in the \Iband bandpass
with WFPC2.

For the \Bj galaxy sample of Frei \etal (1996) with $B_{T}\le$13 mag, we
assume a mean redshift of $\langle z \rangle = 0.007$, based on the mean of
1569 galaxies with measured $V_{gsr}$ km/s ( redshift relative to the galactic
standard of rest) in the RC3 for \Bt $\leq$ 13 mag. The Frei sample was used
to get a rest-frame bandpass as close as possible to that of our HST \Iband
observations of high redshift galaxies, therefore accounting for the
K-correction to first order. We note that \Bj, centered at 4500\AA, may
strictly only be used to simulate the \Iband-band appearance of galaxies in
the range 0.6\cle $z$ \cle1, and that higher redshift modeling will require the
use of large samples of \U images. However, as noted by Odewahn \etal (1996),
the systematic changes across the \Uband, \Bband, \Vband, and \Iband filters
in effective radius and isophotal axis ratio are small (less than 15\%) for
galaxies in the HDF and 53W002 fields (see also figure 1 of Burg \etal 1997). 
Hence, we used the Frei \Bj images to
simulate galaxy images to $z\leq1.6$, where the HST \Iband filter samples
around rest-frame 3000\AA\ . On the basis of $z$ measurements from the CFRS
(Lilly \etal 1995) and the photometric redshift estimates in the HDF (Sawicki
\etal 1997), we adopt a median redshift for our sample of 223 \Iband galaxy
images with \Iband $\leq$ 22 mag of $z_{med}$\cle0.5. The \Bband and \Vband
images of these lower redshift galaxies are used to simulate the appearance 
of more distant galaxies ($z$\cle 1.2) observed in \Iband.


Each of the 63 Frei \Bj galaxy images is used to simulate galaxies at
redshifts of $z=$ 0.3, 0.6, 0.9, and 1.6. Each of the 223 galaxies in the
bright \Iband sample ($z$\cle0.5) is used to simulate galaxies at redshifts of
$z=$ 0.6, 0.8, and 1.2. These simulated images are reduced with exactly
the same MORPHO routines as used in \S~2 for our actual WFPC2 images. 
These simulated measurements are then used to study the change in measured
axis ratio as a function of \SN ratio and linear resolution for every type of
galaxy, using their known ``true'' \ba from the high resolution and high S/N
ground-based images. The \SN ratio of the galaxy image in the distant galaxy
simulation is computed using the integrated signal within an optimally selected
elliptical aperture (measured at an approximate surface brightness of 
$\mu_{I} = 24.5$ \magarc) and the standard deviation of the local sky, whose mode
was determined after a few iterations of outlier rejection. As a measure of
image resolution, $R$, we use the number of pixel elements along the major
axis of this optimal isophotal ellipse. Axis ratio changes are measured with
residuals computed as $(b/a)_{o}-(b/a)_{z}$, where $(b/a)_{o}$ is the
isophotal shape measured in the original local galaxy image, and
$(b/a)_{z}$ is the value measured in the simulated high redshift image. For
subsets of $(b/a)_{o}$, we search for systematic trends with \SN and
resolution to assure that the local comparison set is as much as possible
bias-free.

The results of the \Iband image experiments for galaxies with $(b/a)_{o}$
$\leq$ 0.5 are summarized in Figure~\ref{fig:fig1}. Two major points are clear
from these diagrams: (1) \ba measurements are dominated by accidental errors,
as evidenced by the large scatter in both panels of Figure~\ref{fig:fig1}, and
(2) images with intrinsically small axis ratios are measured systematically
more round (large \ba) as the resolution and/or \SN are decreased. In
Figure~\ref{fig:fig1}, we plot for 176 simulated high $z$ galaxy images a
linear least squares regression line fit for $(b/a)_{o}-(b/a)_{z}$ vs.
$\log$(\SN). Although dominated by a large scatter, the measured slope of $\alpha
= 0.049 \pm 0.014$ is statistically significant at the 3.5$\sigma$ level, and
predicts residuals in the sense expected: faint, low \SN images are measured
to be systematically too round.

In Fig. 1b we show the same fit relative to $\log(R)$ with a fitted regression
of slope $\alpha = 0.067 \pm0.019$, again significant at the 3.5$\sigma$ level, 
and with again the same effect: faint compact images tend to be
measured too round. The residual axis intercepts of these fits predict offsets
of $(b/a)_{o}-(b/a)_{z} = 0.05$ at \SN$=100$ and $R=32$ respectively. Similar
fits using image simulations of galaxies with $(b/a)_{o}$ $\geq$ 0.5 showed no
statistically significant trends with \SN and $R$. {\it Summarizing both results,
intrinsically round galaxies at high redshift are not measured to be
systematically too flat, but intrinsically flat galaxies are measured too
round.} Results obtained with image simulations using the Frei \Bj 
galaxies were consistent with our \Iband-band experiment.
The measured slopes had the same magnitude and sense as for \Iband 
simulations. These results were significant at only the 2$\sigma$ 
level due to a smaller number of input galaxies with $(b/a){o}$ $\leq$ 0.5 
(only 18 such cases). 

As expected, \SN and $R$ are highly correlated
(correlation coefficient of 0.86), and hence it is difficult to determine if
one image property really dominates the effect observed in Figure~\ref{fig:fig1}. 
Independent of this, the important result of these experiments is that we 
can now predict systematic \ba measurement errors on the basis of
easily {\it measured} properties of images. The bulk of the errors resulting
from this effect are accidental, but important systematic trends are derived
from the data in Figure~\ref{fig:fig1} of the form:
 
\begin{equation} (b/a)_{o}-(b/a)_{z} = 0.067 \log R - 0.151 \label{eq:eq5}
\end{equation}

or
 
\begin{equation} (b/a)_{o}-(b/a)_{z} = 0.049 \log S/N - 0.148 \label{eq:eq6}
\end{equation}

Although relatively small, we must consider this systematic effect in \S4 when
we compare local galaxies (observed with high resolution and large \SN) and
high redshift galaxies (poor resolution and low \SN). 
 
\subsection{Consistency of Near and Distant Galaxy Samples}

Since we have chosen galaxies at high redshift on the basis of their apparent
magnitudes in \Iband, we must choose present-day galaxies by the same
criterion in the de-redshifted filter for the expected redshifts of our
distant galaxies.  As such, we choose a bright subset of the
B-band data (\Bt$\leq$13.0 mag) contained in the RC3 for our local sample of
galaxies with measured axial ratios.  For each galaxy, the RC3 gives the
decimal logarithm of the isophotal axis ratio, $R_{25} = (D/d)_{25}$, where
$D$ and $d$ are the major and minor axis lengths measured at 25.0\magarc in
the \Bband (note that the RC3 defines $D_{25}$ in units of 0.1 arcmin). 
As discussed by de Vaucouleurs \etal (1991), the majority of these
axis ratios are based on visual estimates from a variety of sources, but have
been transformed to a self-consistent system via comparisons with detailed
surface photometry for a sample of 608 galaxies. For galaxies fainter than
\Bt\cge13 mag --- the nominal limit of the original Shapley--Ames (1932)
Catalog --- the RC3 galaxies represent a heterogeneous sample which suffers
from a variety of selection effects.

The 1569 galaxies in the RC3 with \Bt$\leq$13.0 mag have other parameters in
the range: 0.0$\leq$\logR$\leq$1.2, \logD$\geq$0.5, and 
--7$\leq$T$\leq$10.0, where $T$ is the morphological Type in the revised Hubble 
system. This galaxy sample is essentially a somewhat enlarged version of the 
original Shapely-Ames Catalog, and should be reasonably complete to \Bt$\leq$13.0 
mag except for dwarf or extremely low surface brightness galaxies.

To test the consistency of the RC3 eyeball axis ratios with our own MORPHO
machine measurements of axis ratios, we compare the MORPHO measurements to the
RC3 measurements for 145 galaxies: 63 galaxies observed in \Bj by Frei \etal
(1996) and 82 UGC galaxies observed in \B by de Jong (1994) using CCD images. 
The automated surface photometry package in MORPHO was used to estimate
isophotal axis ratios for these 145 galaxies at $\mu_{B}=$ 25\magarc.  The
resulting difference, b/a(RC3) -- b/a(MORPHO) is plotted versus b/a(RC3) in
Figure~\ref{fig:fig2}.  A linear least squares fit to only the 63 galaxies of
the Frei \etal data gives a slope of 0.21$\pm0.01$ and an rms scatter
0.07.  The deJong sample does not cover a wide range of b/a (the galaxies were
selected to have b/a $\geq$0.6), but their values of b/a(RC3) -- b/a(MORPHO)
are consistent with those of Frei \etal in the region where their 
ellipticities overlap. Hence, it appears that the eyeball RC3 \ba estimates are
systematically flatter for flat objects, and systematically rounder for round 
objects. 

The origin of this small, but statistically significant trend is unknown at
present, but obviously important for the comparison with HST data.  Based on
this fit we adopt the following correction formula to transform the 1569 RC3
axis ratios to our own MORPHO measurements, so that now for both
the low and the high redshift galaxies we can measure a consistent ellipticity
by machine:

\begin{equation} 
(b/a)_{c} =  0.80 (b/a)_{RC3} + 0.12          \label{eq:eq3}
\end{equation}

\noindent Given the uncertain reason for making such a correction, we will in
\S4 examine how both the original RC3 axial ratio distribution
and the axial ratio distribution for our 1569 local galaxy sample --- when 
transformed to the MORPHO system --- compare to those of high-redshift 
galaxies. 

It is important to note that the linear correction implied by equation
\ref{eq:eq3} is used for convenience, as the present data in this comparison
are too few to derive a formally correct relationship.  The reader will note 
that a correction of this form will deplete the nearly circular apparent 
axis ratios (\ba $\geq$ 0.92), as well as predict unrealistically elongated 
galaxies. This clearly would complicate the derivation of the 
3-D axial ratio on the basis of \fba (not our purpose here), especially in 
light of the triaxility question for early-type systems (Lambas \etal 1992). 
A nonlinear expression is required for such a correction, however 
the lack of highly elongated and circular isophotal shapes in our present sample 
overlapping the RC3 is to small for such a derivation. 
Since the statistical tests discussed in \S 4.2 are most sensitive to changes at 
more intermediate axis ratios, particularly around the median, we feel justified 
in making this correction for the purposes of comparing low and high redshift 
samples. 

\section{APPARENT AXIS RATIO DISTRIBUTIONS - NEAR AND FAR}

\subsection{Subdividing the Distant Sample}

The 16 WFPC2 fields listed in Table~\ref{tab:TAB1} were processed with MORPHO 
to assemble photometric catalogs to the depths given in \S2 on the basis of 
total integration time.  Using the \Iband-band vs. $z$ data of Lilly \etal (1995)
for \Iband $\leq$ 22  and the photometric redshift estimates of Sawicki \etal (1997) 
for \Iband $\leq$ 25, we specified a series of magnitude intervals for drawing 
galaxy samples with progressively higher median redshift and adequate
statistical size.  Unlike the local galaxy samples, fitting isophotal shapes
at a fixed SB level is inadequate. Due to cosmological
effects, the change in observed SB is $\Delta_{\mu} = 10
\log(1+z)$ \magarc. Hence, fitting ellipses at a fixed isophotal level in a
high redshift galaxy gives a measurement of isophotal shape in an
intrinsically brighter region of the galaxy. In the case of an early-type,
bulge-dominated system where the higher surface brightness isophotes often
have rounder shapes than the lower surface brightness outer disky isophotes,
we may expect to see a trend in \ba with surface brightness. The same, or in 
fact the opposite, may be true for the faint Irr galaxies that dominate the
HST images, as we will address in the simulations below.

In order to avoid possible systematic errors in our current analysis, we ran
MORPHO in a mode where multiple ellipse fits are made at progressively fainter
surface brightness levels. These levels are chosen on the basis of \SN in the
local sky-background. For each galaxy, we used the apparent total \Iband
magnitude to estimate a median $z$ for the corresponding magnitude bin. The
cosmological correction to SB, $\Delta_{\mu}$, is then applied to the desired
isophotal level (25.0 \magarc in \Bband was adopted here), and an interpolated
value of the axis ratio at the correct intrinsic surface brightness is 
computed. While not a perfect way to compensate for redshift effects, it is
the best we can do until all the galaxies sampled have individual redshifts
measured for \Iband\cle 26 mag, which is a major challenge even for 8--10
meter telescopes.  The final WFPC2 samples consists of:

\begin{enumerate} 
\item a bright sample of 105 galaxies in the range \Iband$\leq$20 mag with
$\langle$\Iband$\rangle = 19.9$ mag at an estimated median redshift of
$z\approx0.2$ and a mean SB threshold of $\mu_{I} = 23.9$ \magarc;

\item 254 galaxies in the range $20\leq$\Iband$\leq$22 mag with
$\langle$\Iband$\rangle = 21.3$ mag at an estimated median redshift of
$z\approx0.5$ and a mean SB threshold of $\mu_{I} = 24.6$ \magarc;

\item 210 galaxies in the range $22\leq$\Iband$\leq$24 mag with
$\langle$\Iband$\rangle = 22.8$ mag at an estimated median redshift of
$z\approx0.8$ and a mean SB threshold of $\mu_{I} = 25.5$ \magarc; and 

\item a faint sample of 296 galaxies in the range $23\leq$\Iband$\leq$25 mag
with $\langle$\Iband$\rangle = 24.2$ mag at an estimated median redshift of
$z\approx1.0$ and a mean SB threshold of $\mu_{I} = 26.0$ \magarc. 
\end{enumerate}

The differential frequency distribution of apparent axis ratios for each
magnitude selected sample was computed using fixed bins of 0.05 in \ba. The
results are plotted in Figures~\ref{fig:fig3} through \ref{fig:fig5}, where
the WFPC2 samples are represented as open circles connected by thin dotted
lines.

\subsection{Comparing RC3 and Distant Samples}

In accounting for systematic errors in both the distant and nearby galaxy
samples, we have chosen to model how RC3 galaxies would appear at the median
redshift for each distant galaxy sample, including measured \SN and linear
resolution effects (\S3.1).  Various methods are used to simulate the
corresponding \fba distribution we expect to observe for the RC3 sample
at a given median redshift {\it if} the galaxy population mix at that epoch
were the same as our local RC3 sample. The apparent magnitude of a galaxy in
the RC3 sample is estimated for \Ho=75 km/sec/Mpc and \qo=0.5 using:

\begin{equation}
m-M = 42.38 - 5\log h_{0} + 5\log z + 1.086 (1-q_{0}) z + \Delta m_{K} + \Delta m_{L} 
\label{eq:eq4} 
\end{equation} 

\noindent where $\log h_{0} =$\Ho$/100$, and M is the absolute magnitude of the galaxy
computed using \Bt and \Vgsr (from RC3). Type dependent K-corrections, 
$\Delta m_{K}$, were computed from Figure 1 of Lilly \etal (1995) 
and applied to the distance modulus of equation~\ref{eq:eq4}. 
Driver \etal (1995) and Lilly \etal (1995), respectively, note that there is 
substantial evidence of strong luminosity evolution since $z$\cge0.5 for late-type 
galaxies (T \cge Sc) and blue galaxies, respectively. 
The morphological number counts in \Bband by
Odewahn \etal (1996) also show that smaller levels of luminosity (and/or
density) evolution have occurred in \Bj for E+S0 and early-type spiral
populations for z\cle 1.

To account for these possible evolutionary effects on our apparent magnitude
estimates for the redshifted RC3 galaxies, we applied a 
redshift-dependent apparent magnitude increase, $\Delta m_{L}$, for E+S0, Sabc and
Sd/Irr galaxies following the models of Driver \etal (1995b): $\Delta m_{L} = 0$
for all galaxies with $z<0.2$; $\Delta m_{L} = -0.5,-1.0,-1.5$ respectively for
E+S0, Sabc and Sd/Irr galaxies with $0.2 \leq z\leq 0.75$; and $\Delta m_{L} =
-1.0,-1.5,-2.5$ respectively for E+S0, Sabc and Sd/Irr galaxies with $z>0.75$.
These estimates are used only to investigate the possible effects of
luminosity (and/or implied density) evolution on our predicted \fba for
galaxies out to $z$\cle1.2, and we do not suggest that these are 
necessarily the exact amounts of luminosity evolution for each of 
these galaxy classes.

On the basis of the apparent magnitude computed in this manner, each 
redshifted RC3 galaxy was binned by apparent magnitude. 
The resulting magnitude-binned sets of \ba were used to compute 
the observed \fba distributions for a local (\ie RC3-like) population 
seen at high redshift. Each RC3 \ba estimate was corrected to the MORPHO 
system using Equation~\ref{eq:eq3}.  These predicted distributions are plotted 
as solid squares connected by bold solid lines in Figure~\ref{fig:fig3}.  
On the basis of Figure~\ref{fig:fig3}, one might infer a depletion of highly
flattened axis ratios (\ba \cle 0.4) at progressively higher redshifts
($I$\cge 22 mag or $z$\cge0.4) compared to our local population. 

   We have used two methods to compare the WFPC2 \ba distributions with 
those from the $``$redshifted" RC3 data sets. A Kolmogorov-Smirnoff 
probability, $P(KS)$, was computed for each set of distributions to 
assess the likelihood that the two distributions were drawn from the 
same parent population. The $P(KS)$ statistic gives the probability 
that the largest deviation between the two cumulative distributions 
drawn from each data set is due to random fluctuation, and hence a small 
value of $P(KS)$ would indicate that the difference between two data 
sets is statistically significant. We have used the conservative level 
of $P(KS) \leq 0.01$ to confirm that two sets are drawn from a different
parent distribution. Values of $P(KS)$ are tabulated in each \ba set 
compared in the panels of Figures~\ref{fig:fig3} through \ref{fig:fig5}. 
For the brightest and faintest magnitude ranges, the $P(KS)$ probability 
just fails to confirm that \fba for the WFPC2 samples is different from that 
of the RC3 set. The two intermediate ranges, covering $20 \leq $ 
\Iband $\leq 24$, confirm to better than a 99.9\% confidence level that 
the two distributions are drawn from different parent populations. As 
noted by Press \etal (1986), the KS test has the advantage that it is 
a nonparametric test that is independent of data binning, but since all
cumulative distributions are the same at the low and high ends,
it is sensitive primarily to changes in the central region 
of the distribution around the median. We are primarily concerned 
with comparing the low axis ratio end of our high and low redshift 
samples, and hence use a second simple statistic for this case. For 
each \ba set, we compute the ratio, $\phi(b/a)$, of the number 
of galaxies with axis ratio smaller than \ba to the number of 
galaxies with axis ratio larger than \ba. 
Larger values of $\phi_{i}(b/a)$ indicate a higher fraction of flat cases 
(low \ba). To compare each set of distributions, we compute the difference 
between these values, $FD = \phi_{RC3}(0.4) - \phi_{WFPC2}(0.4)$,
to form an index that measures the relative enhancement or depletion of 
flat systems at high redshifts. When $FD$ is large and positive then the
WFPC2 samples must be relatively poor in flat cases relative to RC3. 
The value of $FD$ for each distribution comparison is also plotted in the 
panels of Figures~\ref{fig:fig3} through \ref{fig:fig5}. In Figure~\ref{fig:fig3},
this simple statistic confirms our KS test results in the sense that we do find 
$FD$ to be different and progressively larger in the faint magnitude samples. 

To see a more realistic distribution of axial ratios predicted for the 
RC3 at high redshift, we have to take into account systematic measurement
effects due to degradation of \SN and linear resolution with redshift,
specifically the results of equation~\ref{eq:eq5} and equation~\ref{eq:eq6}
(\S3.1). To do this, the apparent \Iband magnitude estimated for each RC3
galaxy was used to estimate the expected $R$ and \SN that would have been
obtained had the galaxy been observed with an integration time appropriate to
a given data set (\ie HFF, HDF, etc...).  We then applied
equations~\ref{eq:eq5} and \ref{eq:eq6} to the redshifted RC3 distributions
based on these values of $R$ and \SN.

Equations~\ref{eq:eq5} and \ref{eq:eq6} correct the observed \ba values as a
function of log$R$ and $S/N$, with the highest corrections coming for 
the most-edge on galaxies. 
In the present case, this means that the most edge-on galaxies in the RC3 seen
today will have their axial ratios measured rounder with decreasing values of
\SN and $R$.  However, within the errors of the calculations presented in
\S3.1, we were unable to detect a strong systematic change in measured \ba for
galaxies with intrinsic \ba \cge 0.5. In order to avoid a discontinuous jump
in the \ba properties of our corrected high redshift samples, we have applied
a smoothing factor to the correction predicted by equation \ref{eq:eq5}: the full
correction is applied for \ba\cge0.45 with the correction decreasing to 10\%
of that predicted by equation~\ref{eq:eq5} at \ba$=0.8$. The basic properties
of the predicted high redshift \fba distribution were found to be rather
insensitive to the form of this smoothing function. The systematic \ba changes
estimated in this way were applied to all measurements composing the
$``$redshifted RC3" catalog.

The bold-line \fba distributions in Figures~\ref{fig:fig4} are derived from
these ``resolution-corrected'' redshifted RC3 samples. It should be noted that
we applied $R$ and \SN  corrections independently, and found the results to be
unchanged.  For simplicity, we show only the $R$ corrected estimates in our
\fba plots.  The important point to be noted in Figure~\ref{fig:fig4} is
that our apparent dearth of flattened systems at high redshift ($I$\cge 22 mag
or $z$\cge 0.4, see Figure~\ref{fig:fig3}) has generally disappeared after 
applying these corrections. With the exception of the $20\leq$ \Iband $\leq 22$ 
interval, the $P(KS)$ probability indicates no significant difference between 
the sets of low and high redshift distributions. The large jump in $FD$ that 
was observed in the high redshift samples in Figures~\ref{fig:fig3}, which 
indicated a 19\% increase in the relative fraction of flattened 
systems at low redshift, has now dropped to the \cle5\% level.  

Not only do the \SN and $R$ corrections remove many edge-on galaxies from the
high redshift samples, the correction to make the eyeball RC3 axial ratio
distribution consistent with the MORPHO machine measurements does so as well.
If we had made this study comparing just the local RC3 axial ratio
distribution (uncorrected) to that of the distant galaxy samples
(uncorrected), we would have found a very large difference in the numbers of
edge-on galaxies seen today versus those seen at high redshift, with possibly
erroneous conclusions about morphological evolution of galaxies as a
consequence.  However, to do so would be not correct in our opinion.  
Rather, by correcting for detectable systematic effects, we find the present-day 
axial ratio distributions of galaxies --- when moved out to high redshift and 
accounting for all possible systematics as well as we can --- to be very similar 
to those for HST galaxies with 0.3\cle $z$\cle 1.0.


To what extent is the lack of detection of a change in ellipticity in the 
range  0.3\cle $z$\cle 1.0 consistent with what we currently know about the
morphologies of galaxies at these redshifts? As discussed in
Driver \etal (1995), Odewahn \etal (1996), and Abraham \etal (1996), the
galaxy population appears to be increasingly dominated by late-type (types later 
than Sc) at high redshift ($I$\cge 21 mag, $z$ \cge 0.5). The \fba distribution for
E+S0 and early spiral populations are different among the {\it local} galaxies
(Binney \& de Vaucouleurs 1980; Lambas \etal 1992). The E+S0 \fba show a smooth 
increase in the frequency of galaxies with increasing \ba, while the spirals 
show a rather flat distribution with a sharp drop in frequency around
\ba$\approx$0.3. For the very latest types (later than Sd), the \fba distributions
computed in Odewahn (1989) are even more strikingly different from the early-
and mid-spiral samples, with a pronounced redistribution of objects at the
high (=round) \ba end of the distribution. Hence, a population dominated by
late-type spiral systems at high redshifts could produce a relative
overabundance of round objects in \fba with \ba \cge 0.6.

As a test of this possibility, we altered the input catalog of RC3 galaxies to
reflect a population that is dominated by a late-type spiral and irregular
galaxies, as in the deep HST images of Driver \etal (1995a) and Odewahn \etal
(1996). For our original sample of 1569 RC3 galaxies with \Bt$\leq$13, the
type fractional percentages are f(E+S0,Sabc,Sd+Irr) = (32\%, 48\%, 20\%),
respectively. We randomly sampled this catalog to readjust these values to
f(E+S0,Sabc,Sd+Irr) = 10\%, 30\%, and 60\%, to create a sample more in line
with the population fractions implied by the faint WFPC2 morphological studies
mentioned above. This spiral/Irr-rich $``$local" catalog was shifted to the
same high redshift intervals used in Figures~\ref{fig:fig3} (with systematic 
corrections as in Figure~\ref{fig:fig4}), and is shown in Figure~\ref{fig:fig5}.  
As is evident from
comparing Fig. 4 to Fig. 5, we see only small changes in the predicted \fba
functions when compared to the original local RC3 type fractions. The tendency
to produce a more peaked distribution at the high \ba end is only mildly
apparent, and this happens only in the high redshift bins where the observed
late-type fraction is increased due to strong luminosity evolution in the 
late types (see Driver \etal 1995a) and to the loss of early-type galaxies from
the redshifted local samples due to their larger K-corrections. Hence, it
would seem that only an unreasonably extreme alteration of the type fractions
(beyond those allowed by the studies of Driver \etal 1995 and Odewahn \etal
1996) would produce a detectable change in our currently observed WFPC2
samples.

Finally, we think that seeing is believing. Hence, we have put together
montages of galaxies (Figures~\ref{fig:fig6}--~\ref{fig:fig8}) with similar
ranges in b/a including: representative Frei \etal (1996) galaxies seen today
in the B passband; the same Frei \etal galaxies placed at $\rm z \sim 0.6$; and
representative galaxies from our WFPC2 images, spanning a range of \Iband = 14
to 25 mag. As one can see, there is little difference in morphology within
each range of b/a. These figures illustrate that our use of B-band images for
present day galaxies and \Iband images for distant galaxies can produce
systematic effects that are not quantifiable with the present data sets. In
particular, it should be noted that the appearance of the galaxies becomes
more asymmetric in the fainter magnitude intervals. This is most evident in
the low \ba montage of Figure~\ref{fig:fig6}. Whether such changes are due to
genuine morphological variation, or some cosmological effect is still open to
question.  As discussed in Bohlin \etal (1991), Giavalisco \etal (1996) and 
Odewahn \etal (1996), the appearance of some types of galaxies can change 
substantially as a given passband effectively samples bluer or rest-frame $UV$ 
emission at higher redshift.  Even though we have used the \Iband-filter for our 
HST data to get a better match to the rest-frame B-band observations for our local 
sample, this match is not perfect. In the future, we will also make a comparison 
with a U-band survey in progress of a large number of UGC and NGC galaxies with the
VATT telescope (see Burg \etal 1997).


\section{DISCUSSION}

\subsection{Comparison to other HST findings}

  Morphological type fractions have been shown to change systematically
at fainter magnitudes or at higher redshift.  
The rather drastic increase in late-type systems
at z\cge0.5 observed by Driver \etal (1995), Glazebrook \etal (1995), and
Odewahn (1996) is interpreted as evolution among the galaxy populations
at high redshift. From numerous papers referenced in our introduction, it is 
well known that early- and late-type systems display different apparent 
axis ratio distributions locally, and hence it
was surprising that a large change in the fraction of Sd/Irr galaxies at high
redshift would not manifest itself strongly in the observed axis ratio
distributions. On the contrary, as we have shown in the previous section,
only small deviations in the \fba distribution for high redshift galaxy samples are
present compared with local samples. In fact, these small deviations
may be totally explained if we take into account systematic errors in
isophotal parameters measured in images with low linear resolution and
low signal to noise. The observed \fba distribution for faint galaxy samples will also
be affected by two additional factors: weak lensing and the possible
influence of a population of linear "chain galaxies" at high redshift.

The shapes and orientations of distant galaxy images will be distorted from
gravitational lensing by intervening mass distributions. Tyson \etal (1990)
and Kaiser and Squires (1993) have used this effect to probe the mass distributions
in rich clusters in an effort to confirm mass determinations derived from
purely dynamical methods. The samples used in our current study were
drawn from a variety of positions on the sky and are not expected
to be profoundly influenced by foreground rich clusters. The presence
of rich clusters in our current samples would manifest themselves by
increased number counts relative to the field. On the contrary, the
(morphological) number counts derived from the samples used here were found
by Driver \etal (1995) and Odewahn \etal (1996) to be very consistent.

However, the weak distortion of faint galaxy images caused by lensing from
large scale structures (LSS) in the Universe must be considered. Large scale
deviations in the mass distribution inferred from systematic galaxy redshift
surveys by Huchra \etal (1988) and the measurement of fluctuations in
the cosmic microwave background by COBE (Smoot \etal 1992) indicate that large
matter overdensities may be common throughout the Universe, and that lensing
resulting from these inhomogeneities may produce ellipticity changes of up
to 3\% for galaxies as distant as z$\approx$1. A recent measurement of this 
weak distortion by LSS is presented by Mould \etal (1994), in a ground-based 
study of 4363 galaxies to a limit of r$\approx$26 mag. 
They find a weak lensing signal significant at the
1$\sigma$ level in the form of a mean position angle polarization of 1\% over
that expected for a random distribution. The detection of galaxy-galaxy lensing
has been studied by Tyson \etal (1985) in a sample of 47000 galaxy images
in the range 22.5$\leq$ \Bj $\leq$ 23.5 with no statistically significant
signal above that expected for a purely random distribution.  More recent
CCD-based studies by Brainerd \etal (1996) use 3202 galaxy pairs in the
range 23 $\leq$r $\leq$ 24 mag to detect a lensing signal in the form of a
position angle polarization of 1.1\% $\pm$0.6\%. They report that an
ellipticity of approximately 2-3\%, similar to the observed polarization
effect, will be induced in faint galaxy images with angular separations
of $\theta \approx 3''$ from bright galaxies. In the course of this
ground-based study, they considered the ellipticity distribution of
galaxies from their faint samples, and determine that the
ellipticity distributions of these objects are consistent with
\fba drawn from brighter, more local samples. Finally, the
presence of a large amount of image distortion by LSS should be
accompanied by a large amount of multiple imaging, an effect
which is not observed commonly in well-studied distant
radio galaxy samples (Blandford, private communication). The overall
conclusion from these studies is that weak gravitationally lensing may
certainly influence the observed \fba of faint galaxies by $``$stretching"
the isophotal image contours. However, recent redshift surveys by
Lilly \etal (1995) and Ellis \etal (1996) suggest that the median 
redshift of the galaxy sample used here is $z$\cle1, and thus the overall
image distortions due to lensing will produce a rather small 1-2\%
systematic effect.

Separately, a population of faint blue galaxies with very narrow shapes and lumpy
morphologies was reported by Cowie \etal (1996) and termed ``chain'' galaxies,
a possibly new morphological class of galaxy common during younger epochs of
the Universe. A truly linear structure, such as the chain galaxies are thought
to be, will present an isophotal shape having a small value of b/a in nearly
any spatial orientation. Hence, the presence of a large population of such
galaxies in our faint samples could skew our observed \fba towards the
low b/a end.  The presence of such a new morphological population has been
refuted by Dalcanton and Shectman (1996), who claim that these galaxies do
indeed exist at low redshift and represent edge-on views of low surface
brightness late-type (T \cge Sd) disk galaxies. Using a low-redshift sample of
low surface brightness galaxies (central disk surface brightness $\mu \geq$
21.3 \magarc in the r band), Dalcanton \& Shectman show that the \fba of this chain
galaxy sample is consistent with that of a population of thin disk galaxies
viewed with random orientations.  The galaxies with the the smallest \ba
display morphological similarities with the ``chain'' galaxies, especially in
the far-$UV$. 

  The UIT images by O'Connell \etal (1997) and Marcum \etal (1997)
of ordinary nearby disk galaxies seen edge-on resemble those of 
high redshift ``chain galaxies'', so that the clumpy rest-frame $UV$-morphology 
of such objects at high redshift may be classified as a different or new class 
of object, while in reality they may not be. Our observations are 
consistent with this latter view. However, Pascarelle \etal (1996) note that 
the Cowie \etal (1996) chain galaxies are a clear minority of the faint 
blue galaxy population in deep WFPC2 images. The faint chain galaxies often do 
show clear ``dumbbell'' structure with most of the light in clumps at both outer 
ends --- unlike many of the UIT images of nearby edge-on galaxies --- so that these 
objects might be subgalactic sized clumps seen directly after or during their 
first mergers, which would trigger subsequent starbursts and form disks from 
gaseous material that settles into rotation during and after the merger.
Hence, they could be short-lived ``chain-like'' objects, but their fraction of 
the total population is not large enough to significantly affect our 
\ba distributions. 


\subsection{Comparison to the Hierarchical Clustering/Merging Galaxy Formation Scenario} 

Consider what we might expect to see at high redshift from galaxies that form
via the hierarchical cluster/merger hypothesis (e.g. Gott \& Rees 1975). 
Peebles (1993) defines lookback time to be the time measured backwards from
the present time, in other words, since $z=0$. 
Hence, larger values of lookback time correspond to earlier cosmic epochs
(and larger redshift) in the Universe.  The familiar galaxy types we see today are 
thought to first start out in the form of $10^7 - 10^8$ M$_\odot$ clumps of gas and
newly formed stars. We will first see such pre-galactic clumps when they give
off sufficient radiation to be detected, at a lookback time that we call $\rm t_f$.
The Pascarelle \etal (1996) cluster of 18 subgalactic sized objects at z$\simeq$2.4 
is an example of this process. We stress, however, that in hierarchical CDM models 
the formation of these clumps does not have to be coeval, and will in fact 
likely happen over a wide range of epochs (z$\simeq$1--4). 
When these clumps start to combine and 
form ever larger clumps, at a smaller lookback time $\rm t_m$ ($<\rm t_f$), we will 
begin to see the merged clumps as galaxies of morphologies that are familiar to us 
today: E/S0's, Sabc's, Sd/ Irr's.
 
At lookback times smaller than $\rm t_m$ (i.e., after the merging epoch began),
hierarchical models predict that we will see galaxies with a range of
morphological types not dissimilar to what we see today, but with a galaxy mix
that is not necessarily equal to the local one.  
What we will {\it not} know --- if hierarchical clustering is correct --- is 
what a particular galaxy at a lookback time smaller than $\rm t_m$ will look 
like today.  Will it simply passively evolve? Or will it merge with its 
nearest neighbors, turning (temporarily) into another type of galaxy 
altogether? In other words, in a hierarchical merging scenario, while the 
galaxy population mix may not change in a statistical manner for lookback 
times less than $t_m$, one cannot necessarily uniquely assign galaxy of type X 
at high redshift with galaxy of type X today.
 
The lack of change in the distribution of edge-on systems with ellipticity
that we observe is consistent with hierarchical clustering in the following
sense.  We know from the distribution of Hubble types today that
bulge-dominated galaxies comprise the minority of giant galaxies (less than
30\%, see Dressler 1980).  If all galaxies initially are formed
as disk-type systems, with subsequent merging producing pronounced spiral
bulges and E/S0 galaxies, then we would expect a steady migration from
galaxies with low b/a to those with high b/a.
While such a trend is suggested in the present comparison, the observed result
is barely of one-sigma significance.  The problem is observational:  It is
difficult to accurately measure the ellipticities of galaxies with
b/a $>$ 0.85.  Errors in photometry will always scatter ellipticities away from
b/a = 1.0 (since \ba cannot be larger than 1), always depopulating the 
highest bin in \fba. As discussed in the introduction and in \S 3.2, our 
systematic correction of the RC3 axis ratios, as well as uncertainties 
related to the triaxiality of early-type galaxies, makes treatment of 
the high \ba systems problematic.  
 
If there is one place in comparing the ellipticity distributions of
nearby and distant galaxies to hide evolutionary effects, it is among galaxies
with b/a $\ge$ 0.85.  Yet, this is where many merger products are likely to
end up in terms of their intrinsic axial ratios.  What we conclude is that nature
apparently conspires to have systematic effects (observational ones and those
due to the redshift) on the ellipticity distributions to essentially cancel
the effects expected from the evolutionary change in galaxy morphology.
If the hierarchical clustering/merging picture is correct, then the kind of 
ellipticity test that we have done in this paper is in essence a 
null test. If we had seen a  difference at high ellipticities, it would 
have been a surprising result.  By seeing none, we are consistent with the 
standard hierarchical clustering/merger model.
 
We believe that there is enough evidence now to give a reasonable estimate to
the redshift corresponding to a lookback time of $\rm t_m$. Pascarelle \etal
(1996) have shown that at least in one direction at z$=$2.39, we still find 
the subclumps that have generally not yet merged into morphologies recognizable 
today. If we make the reasonable estimate that the high redshift galaxies in 
our ellipticity sample are outnumbered by those at z \cle 1 (see Figure 2 
of Sawicki \etal 1997), we conclude that whatever changes in morphology 
that have occurred since z$\leq$1 must have preferentially occurred
among round-appearing galaxies - i.e., E, S0 and bulge-dominated spirals.
Mergers of systems we would recognize today as galaxies classifiable by
the Hubble scheme have therefore been taking place at least as far back as 
z$\simeq$1, but likely no farther back than z$\simeq$2.4. This would argue that 
1.0\cle $z_{t_m}$\cle 2.5, with the most likely time being 
1.5\cle$z_{t_m }$\cle 2.0, consistent with the peak in the star-formation 
rate in the universe (Madau \etal 1996).
 
If this view of hierarchical clustering is correct, then
understanding how galaxies evolve with time will not be a simple matter of
estimating galaxy morphologies and colors at all lookback times. Rather, if
what we see at lookback times smaller than  $t_m$, including the present-day, 
is the current state of mergers, we will need more detailed information about how
groupings of galaxies change with time.  That information can only come with
measurements of redshifts of very faint galaxies, which is a major challenge that
the new generation of 8--10 meter telescopes face if we are to truly understand
galaxy evolution.
 
 

We acknowledge support from HST grants GO.5308.01.93A, GO.5985.01.94A,
GO.2684.03.94A, and AR.6385.01.95A. SCO wishes to acknowledge a very 
useful conversation with R. Blandford and D. Hogg. 

\section*{REFERENCES} 
\begin{description} 

\item Abraham, R., Tanvir, N.R., Santiago, B., Ellis, R. S., Glazebrook, K. G.
\& van den Bergh, S., 1996, \mnras, 279, L47

\item Bertin, E., \& Arnouts, K., 1996, in preparation 

\item Bohlin, R. C., \etal 1991, \apj, 368, 12


\item Brainerd, T., Blandford, R. D., \& Smail, I. 1996, \apj, 466, 623

\item Binney, J., \& de Vaucouleurs, G. 1980, \mnras, 194, 679

\item Burg, C., Windhorst, R., Odewahn, S., de Jong, R., \& Frogel, J.  1997 in 
    Proceedings of the Maryland Workshop on ``The Ultraviolet Universe at Low 
    and High Redshift: Probing the Progress of Galaxy Evolution'', Eds. M.  
    Fanelli, \& W. Waller, AIP Conf. Ser., in press

\item Cowie, L. L., Songaila, A., Hu, E. M., \& Cohen, J. G. 1996, \aj, 112,
839

\item Dalcanton, J., \& Shectman, S. A., 1996, \apjl, 465, L9

\item de Vaucouleurs, G., de Vaucouleurs, A., Corwin, H.G., Buta, R., Paturel,
G. \& Fougu\'{e}, P. 1991, The Third Reference Catalog of Bright Galaxies,
Springer Verlag: New York, (RC3)

\item de Vaucouleurs, G. \& Pence 1973, BAAS, 5, 466

\item de Vaucouleurs, G. 1959, {\it Handbuch der Physik}, 53, 311

\item de Jong, R. S., \& van der Kruit, P. C. 1994, \aaps, 106, 451.

\item Dressler, A. 1980, \apj, 236, 351 

\item Driver, S. P., Windhorst, R. A., Ostrander, E. J., Keel, W. C.,
Griffiths, R. E. \& Ratnatunga, K. U. 1995a, \apjl, 449, L23 (D95a)

\item Driver, S. P., Windhorst, R. A., \& Griffiths, R. E. 1995b, \apj, 
  453, 48 (D95b)

\item Ellis, R., Colless, M., Broadhurst, M., Heyl, J., \& Glazebrook, K.
1996, \mnras, 280, 235

\item Fasano, G. \& Vio, R. 1991, \mnras, 249, 629

\item Ferguson, H.C. \& Sandage, A. 1989, \apjl, 346, L53

\item Frei, Z. \etal 1996, \aj, 111, 174 

\item Giavalisco, M., Livio, M., Bohlin, R., Macchetto, D. F., \& Stecher, T.
P. 1996, \aj, 112, 364

\item Glazebrook, K., Ellis, R. E. Santiago, B., \& Griffiths, R. E. 1995,
\mnras, 275, L19

\item Gott, J. R., \& Rees, M. R. 1975, A\&A 15, 235


\item Hill, J. K. \etal 1992, \apjl, 395, L37  

\item Holtzman, J. A., \etal 1995, \pasp, 107, 1065

\item Hubble, E. 1926, \apj, 64, 321

\item Huchra, J., Geller, M., De Lapparent, V., \& Burg, R., 1988, IAU
Symposium 130, Large-Scale Structures of the Universe, ed. J. Audouze, M.
Pelleto, and A. Szalay, Kluwer, 105

\item Im, M. \etal 1995, \apjl, 445, L15



\item Kaiser, N. \& Squires, G. 1993, \apj, 404, 441

\item Lambas, D. G., Maddox, S.J., \& Loveday, J. 1992, \mnras, 258, 404

\item Lilly, S. J., Tresse, L., Hammer, F., Crampton, D., \& LeFevre, O. 1995,
\apj, 455, 108


\item Madau, P. \etal 1996, \mnras, 283, 1388 

\item Marcum, P. \etal 1997 in Proceedings of the Maryland Workshop on 
  ``The Ultraviolet Universe at Low and High Redshift: Probing the Progress of 
    Galaxy Evolution'', Eds. M.  Fanelli, \& W. Waller, AIP Conf. Ser., in press


\item Mould, J., Blandford, R., Villumsen, J., Brainerd, T., Smail, I., Small,
T., \& Kells, W. 1994, \mnras, 271, 31

\item Mutz, S. B., \etal 1994, \apjl, 434, L55

\item Noerdlinger, A. 1979, \apj, 234, 802

\item O'Connell, R. \etal 1997 in Proceedings of the Maryland Workshop on 
  ``The Ultraviolet Universe at Low and High Redshift: Probing the Progress of 
    Galaxy Evolution'', Eds. M.  Fanelli, \& W. Waller, AIP Conf. Ser., in press

\item Odewahn, S.C. 1997, Nonlinear Signal and Image Analysis, Annals of the New 
      York Academy of Sciences, 188, 184 

\item Odewahn, S.C., Windhorst, R. A., Driver, S. P., \& Keel, W. C. 1996,
\apjl, 472, L13

\item Odewahn, S. C, \& Aldering, G. 1995, \aj, 110, 2009

\item Odewahn, S. C. 1995, \pasp, 107, 770

\item Odewahn, S.C. 1989, Ph.D. thesis, Univ. of Texas

\item Pascarelle, S. M., Windhorst, R. A., Keel, W. C., \& Odewahn, S. C. 1996,
      \nat, 383, 45

\item Peebles, P. J. E. 1993, ``Principles of Physical Cosmology'', (Princeton 
University Press), Princeton, New Jersey

\item Press, W. H., Flannery, B. P., Teukolsky, S. A., and Vetterling, W. T. 1986, 
      Numerical Recipes, (Cambridge University Press), 472

\item Ryden, B. S. \& Terndrup, D. M. 1994, \apj, 425, 43

\item Sandage, A., Freeman, K. C., \& Stokes, N. 1970, \apj, 160, 831


\item Sawicki, M. J., Lin, H., \& Yee, H. 1997, \aj, 113, 1

\item Smoot, G. F., Bennett, C. L., Kogut, A., \& Wright, E. L., 1992, \apj,
396, L7

\item Tyson, J. A. 1985, \nat, 316, 799

\item Tyson, J. A., Valdes, F.,  \& Wenk, R. A. 1990, \apj, 349, L1 

\item Williams, R.E. \etal 1996, \aj, 112, 1335  
 
\item Windhorst, R. A., Keel, W. C., \& Pascarelle, S. M. 1997, \apjl, submitted

\item Windhorst, R. A., Burstein, D., Mathis, D. F., Neuschaefer, L. W., Bertola, F., 
      Buson, L., Matthews, K., Barthel, P. D., \& Chambers, K. C. 1991, \apj, 380, 362 

\item Weedman, D. W., 1986, in {\it Quasar Astronomy} (Cambridge: Cambridge
University Press) 71

\end{description} 

\baselineskip=18pt 

\clearpage

\begin{figure}
\caption{ Using 176 simulated high $z$ images based on observed \Iband-band
WFPC2 images with \I$\leq$22 (and assumed $z$\cle0.5) and intrinsic axis ratio
\ba$\leq$0.5, we plot the residual axis ratio, $(b/a)_{o}-(b/a)_{z}$, as a
function of signal-to-noise ratio ($\log$($S/N$) in panel a (left) and image
resolution $\log(R)$ in panel b (right). Both of these tests predict residuals
in the sense expected: faint, low \SN compact images are measured to be
systematically too round.
}
\label{fig:fig1}
\end{figure}

\begin{figure}
\caption{Comparison between the eyeball RC3 \ba values derived from \logR and
those derived from the automated MORPHO package. Shown are 63 galaxies
observed in \Bj by Frei \etal (1996; open circles) and 82 galaxies observed in
\Bccd by de Jong (1996; open squares). A linear least squares fit to the
(eyeball--machine) residuals gives a slope of 0.21$\pm0.01$ and an r.m.s.
scatter about the best fit (dashed) line of 0.07. The fit was
made only with the Frei data which cover an adequate range in b/a, but the
de Jong sample (by selection has \ba$\geq$0.6), were plotted to show
consistency with the derived fit. Based on this fit, we adopted the correction
formula in the text (equation \ref{eq:eq3} to transform the 1569 RC3 
(eyeball) axis ratios to the system defined by the MORPHO automated 
ellipse fitting software used to reduce all our WFPC2 faint galaxy images.
}
\label{fig:fig2}
\end{figure}

\begin{figure}
\caption{The differential frequency distributions of apparent axis ratio, \ba,
for four \Iband-band ranges are plotted using open circles connected by thin
dashed lines. The {\it local} RC3 galaxy \ba distribution is plotted in each
panel using filled squares connected by bold lines. The RC3 data were
redshifted assuming \Ho=75 km/sec/Mpc, \qo=0.5, plus K-corrections and
luminosity evolution estimates discussed in the text. No systematic
corrections for $R$ or \SN effects have been applied to the RC3 data in this
figure.
}
\label{fig:fig3}
\end{figure}

\begin{figure}
\caption{ As in Figure~\ref{fig:fig3}, but systematic corrections for
$R$esolution effects have been applied to the RC3 \ba estimates.  No
statistically significant difference is seen between the RC3 sample moved out
ot high redshift and the observed WFPC2 \Iband-band samples, even for galaxies
with low \ba values.
}
\label{fig:fig4}
\end{figure}

\begin{figure}
\caption{As in Figure~\ref{fig:fig3}, but the local Type fractions in the RC3
for E+S0, Sabc, and Sd+Irr galaxies have been altered from 32\%, 48\%, and
18\%, respectively, to a spiral-rich sample of 10\%, 30\%, and 60\% at
$z\simeq 1$. Systematic corrections for $R$ effects have been applied to the
RC3 \ba estimates, as discussed in the text. Note that with these corrections,
the higher and lower redshift samples are virtually indistinguishable for
$I$\cge22 mag ($z$\cge 0.5).
}
\label{fig:fig5}
\end{figure}

\begin{figure}
\caption{ A montage of galaxies with $0.0\leq$ \ba $\leq0.4$ selected at
random from the ground-based and HST galaxy samples used in this work. The top
row consists of 8 galaxies (observed in \Bj) by Frei \etal 1996, and the
second shows the appearance of these same galaxies at $z$=0.6,as modeled by
MORPHO. In the first row, images are shown at ground-based resolution with 
each postage stamp image having a linear size of 167$''$. In all remaining 
rows, a WFPC2 resolution (0.0996$''$/pix) is used with each stamp having 
a linear dimension of 6.2$''$. The remaining 4 rows
consist of WFPC2 images in \Iband roughly corresponding to the four magnitude
ranges constituting our faint samples: 14$\leq$ \Iband $\leq$20, 20$\leq$
\Iband $\leq$22, 22$\leq$ \Iband $\leq$24, 24$\leq$ \Iband $\leq$25. Each
image is normalized to a linear grey-scale from $-3\sigma$ to $+20\sigma$,
where $\sigma$ is the noise measured in the local background. Morphological
features such as spiral arms or strong dust lanes are not prominently seen in
the $z\geq0.4$ rows of WFPC2 images, although such features are still clearly
visible in the \Bj images modeled for $z=0.6$. The high redshift systems 
($z\geq0.8$) do appear to display a higher degree of image asymmetry 
along the major axis and more clumpiness relative to the lower redshift systems.
}
\label{fig:fig6}
\end{figure}

\begin{figure}
\caption{ As in Figure~\ref{fig:fig6}, but for galaxies with $0.6\leq$ \ba
$\leq0.75$. As in the \ba$\leq$0.4 cases of Figure~\ref{fig:fig6}, it appears
that prominent spiral arms are not seen in the $z\geq0.8$ rows of WFPC2
images, and the degree of image clumpiness and asymmetry is higher compared to
their low redshift counterparts.
}
\label{fig:fig7}
\end{figure}

\begin{figure}
\caption{ As in Figure~\ref{fig:fig6}, but for galaxies with $0.8\leq$ \ba
$\leq1.0$. Again no prominent spiral structure is visible for $z\geq0.8$, but
now the impression of increased global image asymmetry is not as strong for the
rounder galaxies.
}
\label{fig:fig8}
\end{figure}

\begin{deluxetable}{l c c c c c}
\footnotesize
\tablecaption{WFPC2 Fields Analyzed in F814W}
\tablewidth{0pt}
\tablehead{
\colhead{Name}   &  \colhead{ RA (2000) DEC }  &
\colhead{l,b}   &  \colhead{$A_{B}$}  &
\colhead{T(sec)} &  \colhead{ $N_{images}$ }
}
\startdata
HDF       & 12:36:49.59  +62:12:54.0 & 125.9 +54.8 &  0.00 & 123600  &  49 \nl
53W002    & 17:14:15.47  +50:15:28.3 &  77.0 +35.8 &  0.03 & 1700    &  17 \nl
HFF\_OW   & 12:36:21.16  +62:14:27.2 & 126.0 +54.8 &  0.00 &  2500   &  3  \nl
HFF\_SW   & 12:36:33.91  +62:10:54.2 & 125.9 +54.9 &  0.00 &  2500   &  3  \nl
HFF\_IW   & 12:36:35.77  +62:13:40.9 & 125.9 +54.8 &  0.00 &  5300   &  4  \nl
HFF\_SE   & 12:36:47.77  +62:10:19.3 & 125.9 +54.9 &  0.00 &  2500   &  3  \nl
HFF\_NW   & 12:36:50.81  +62:15:41.9 & 125.9 +54.8 &  0.00 &  2500   &  3  \nl
HFF\_IE   & 12:37:03.86  +62:12:15.2 & 125.8 +54.8 &  0.00 &  5300   &  4  \nl
HFF\_NE   & 12:37:05.85  +62:15:06.7 & 125.8 +54.8 &  0.00 &  2500   &  3  \nl
HFF\_OE   & 12:37:17.92  +62:11:33.4 & 125.8 +54.9 &  0.00 &  3000   &  3  \nl
DWG\_UEH0 & 00:53:23.57  +12:33:50.3 & 123.7 -50.3 &  0.18 &  2100   &  3  \nl
DWG\_UIM0 & 03:55:31.91  +09:43:38.5 & 179.8 -32.1 &  0.46 &  866    &  6  \nl
DWG\_UOP0 & 07:50:47.39  +14:40:42.7 & 206.1 +19.6 &  0.06 &  2100   &  2  \nl
DWG\_UY40 & 14:34:48.54  +25:08:11.1 & 033.9 +66.8 &  0.04 &  1000   &  6  \nl
DWG\_UX40 & 15:19:41.22  +23:52:04.3 & 035.8 +56.5 &  0.14 &  1875   &  4  \nl
DWG\_USA2 & 17:12:23.17  +33:35:48.5 & 056.7 +34.3 &  0.12 &  2100   &  3  \nl
\enddata
\label{tab:TAB1}
\end{deluxetable}

\end{document}